\documentclass[twocolumn,english,aps,prl,superscriptaddress,showpacs,floatfix]{revtex4} 
\usepackage{graphicx}
\usepackage{amsmath,amssymb,amsfonts}
\usepackage{natbib}
\usepackage{babel}
\usepackage{lipsum}
\usepackage{tikz}
\usetikzlibrary{shapes,arrows}
\usepackage{dcolumn}
\usepackage{bm}
\usepackage[makeroom]{cancel}
\usepackage{changes}
\usepackage{enumerate}

\newcommand{\fref}[1]{Fig.\hspace{0.025in}\ref{#1}}

\newcommand{\eref}[1]{Eq.\hspace{0.025in}(\ref{#1})}

\DeclareMathAlphabet\mathbfcal{OMS}{cmsy}{b}{n}

\begin{document}

\preprint{APS/123-QED}

\title{Proposal for a clumsiness-free test of macroscopic realism}

\author{Devashish Pandey}
\affiliation{Department of Electronic Engineering, Universitat Aut\`onoma de Barcelona, 08193 Bellaterra, Spain}%
\author{Xavier Oriols}
\affiliation{Department of Electronic Engineering, Universitat Aut\`onoma de Barcelona, 08193 Bellaterra, Spain}%
\author{Guillermo Albareda}
\affiliation{Max Planck Institute for the Structure and Dynamics of Matter, 22761 Hamburg, Germany}%
\affiliation{Institute of Theoretical and Computational Chemistry, Universitat de Barcelona, 08028 Barcelona, Spain}
\email{guillermo.albareda@mpsd.mpg.de,xavier.oriols@uab.cat}


\date{\today}

\begin{abstract}
We propose a test of macrorealism that exploits the contextuality of two-time correlation functions to escape the so-called ``clumsiness loophole'' that plagues Leggett-Garg inequalities.
The non-contextuality of reduced joint probability distributions is proven to be an unequivocal criterion to guarantee that measurements are carried out in the ideally-weak measurement regime of a class of generalized von Neumann measurements.
In this regime, testing the so-called ``no-signaling in time'' condition allows to uncontextually ascertain whether a property of a given system is macrorealistic or non-macrorealistic. Interestingly, the resulting protocol allows for tests of macrorealism in situations where Leggett-Garg inequalities and ideal negative measurement cannot be used at all.
\end{abstract}

\pacs{Valid PACS appear here}
\maketitle

Ever since the birth of quantum mechanics, theoretical works have deepened our understanding of its conceptual and mathematical structure. In this respect, any list of highlights should definitively include Bell's~\cite{bell2004speakable} and Leggett-Garg~\cite{leggett1985} inequalities to disprove local~\cite{einstein1935can,bell1964einstein} and macroscopic realism~\cite{leggett1988experimental,leggett2008realism} respectively. 
In contrast to the violation of local realism~\cite{aspect1999bell,brunner2014bell,reid2009colloquium}, however, an inarguable violation of macrorealism has remained elusive to date~\cite{wilde2012addressing,emary2013}.
The reason is that whilst special relativity can be used to close the ``communication loophole'' in a Bell test of local realism~\cite{clauser1978bell,kwiat1994proposal,huelga1995loophole,rosenfeld2009towards}, no such defence exists for a Leggett-Garg test of macrorealism. 

Macrorealism does not assert that it is impossible to affect a physical system by measurement and therefore a violation of Leggett-Garg inequalities can only be a proof that system's properties are either (i) non-macrorealistic or (ii) macrorealistic but subjected to a measurement technique that happens to disturb the system. 
This problem is known as the ``clumsiness loophole''~\cite{wilde2012addressing}, and can always be exploited to refute the implications of a Leggett-Garg test of macrorealism. 
While a number of works have addressed this problem by making the explanation of Leggett-Garg inequalities violations in terms of experimental clumsiness so contrived as to be doubtful~\cite{kwiat1994proposal,knee2012violation}, whether a loophole-free Leggett-Garg protocol can be constructed remains an open question~\cite{emary2013}. 

In this Letter we propose a clumsiness-free test of macrorealism that relies on the notion of contextuality introduced by Bell~\cite{bell2004speakable}, Kochen and Specker~\cite{Kochen1975}, which is known to yield observable effects at the level of time-correlation functions~\cite{anastopoulos2006classical,dressel2012contextual,dressel2010contextual}. 
Measuring an observable A at time $t$ and correlating the outcome, $y_A(t)$, with the measured value of B, $y_B(\tau)$, at a later time $\tau \geq t$, represents an unequivocal way of representing the dynamics of classical systems in terms of joint probabilities, i.e., $P(y_A,y_B)
    \leftrightarrow
    \textit{system dynamics}$.
In quantum mechanics, however, the unavoidable backaction of the measurement process~\cite{braginsky1995quantum,dicke1981interaction} precludes such a clear-cut connection. 
Even using the best technological means, different measurement schemes, $\{\sigma_A\}$, can yield different probability distributions, i.e., $P(y_A,y_B)_{\sigma_A}     \leftrightarrow
    \textit{system+apparatus dynamics}$. 
This property of quantum mechanics can result in contradictions among tests of macrorealism that are based on different experimental set-ups. 

To avoid contextuality, we will first identify the \textit{ideally-weak measurement regime}: the regime where, for pure states, two-time correlation functions distinctively unravel either (i) the expectation value of two-time Heisenberg operators for non-macrorealistic properties or (ii) the product of the expectation values of two independent events for macrorealistic ones. We will then show that this regime can be experimentally identified by witnessing the non-contextuality of a reduced probability distribution. Finally, we will prove that for general (mixed) states, assessing the so-called ``no-signaling in time'' criterion~\cite{kofler2013condition,li2012witnessing} under ideally-weak measurement conditions makes it possible to unambiguously distinguish between macrorealistic and non-macrorealistic properties.

We consider a generalized von Neumann measurement~\cite{von2018,jacobs2014}, where the expectation value of a property A, associated to the operator 
$\hat A = \sum_i a_i |a_i\rangle \langle a_i |$ (with $a_i$ and $| a_i \rangle$ being the corresponding eigen-values and -states), of a quantum system $|\psi(t)\rangle$ is determined by repeatedly reading-out the pointer position of the meter over a large ensemble of identically prepared experiments:
\begin{equation}\label{exp_y}
 \langle y_A(t) \rangle = \int_{-\infty}^{\infty} dy_A y_A  P(y_A),
\end{equation}
where $P(y_A)$ is the probability of finding a value $y_A$ of the pointer position at time $t$. According to Born's rule, $P(y_A)$ can be expressed in terms of the system degrees of freedom as $P(y_A)=|\psi_A(t)|^2$, where 
\begin{equation}\label{def_kstate}
  |\psi_A(t) \rangle = \sum_{i} \Omega_{y_A-\lambda a_i} c_i |a_i\rangle
\end{equation}
is the state of the system right after measuring $y_A$ at time $t$~\cite{AppA}. 
In \eref{def_kstate} we have defined the coefficients $c_i = \langle a_i|\psi(t)\rangle$, and the displaced (by an amount $\lambda a_i$) wavepacket of the pointer, $\Omega_{y_A-\lambda a_i}$, with $\lambda$ being a macroscopic parameter with units of $[L][A]^{-1}$ that hereafter is assumed to be $\lambda = 1$~\cite{misc3}. In order to ensure that \eref{exp_y} always yields the correct expectation value $\langle y_A(t) \rangle = \langle \hat A \rangle$, it is enough to make the pointer wavepacket to be well normalized and obeying  $\int y_A |\Omega_{y_A-a_i}|^{2} dy_A = a_i$ ~\cite{aharonov1991complete,kofman2012,jacobs2014}.

A second, subsequent, measurement of a property B, associated to the operator $\hat B = \sum_i b_i |b_i\rangle \langle b_i |$ (with $b_i$ and $| b_i \rangle$ being the corresponding eigen-values and -states) can be easily accommodated into the above scheme by simply reading-out the pointer position of a second measuring apparatus at time $\tau\geq t$. The two-time correlation function $\langle y_A(t)y_B(\tau)\rangle$ can be then evaluated as:
\begin{eqnarray}\label{two}
 \langle y_A(t)y_B(\tau)\rangle = \int_{-\infty}^{\infty}  \int_{-\infty}^{\infty} dy_A dy_B \; y_A y_B P(y_A,y_B),
\end{eqnarray}
where $P(y_A,y_B)$ is the joint probability of reading-out the values $y_A$ and $y_B$ at times $t$ and $\tau$ respectively.
Using Born's rule, this probability can be written as $P(y_A,y_B) =  |\psi_{A,B}(\tau)|^2$, where
\begin{equation}\label{def_kwstate}
 |\psi_{A,B}(\tau)\rangle = \sum_{i,j} \Omega_{y_A-a_i}\Omega_{y_B-b_j} c_{i}c_{j,i}|b_{j}\rangle 
\end{equation}
is the state of the system right after the two-time measurement process~\cite{AppB}. In \eref{def_kwstate} we have defined $c_{j,i}  = \langle b_j | \hat U_\tau|a_i\rangle$, and $\hat U_\tau = \exp(i\hat H\tau/\hbar)$ describes the unitary evolution of the system between the two measurements.

Without the loss of generality, we can now restrict the meter wavepacket to be represented by a gaussian Krauss operator~\cite{kraus1983states,wiseman2009}, i.e., 
  $\Omega_{y - a_j} = \mathcal{A} \exp{\left[-{(y-a_j)^2}/{4\sigma^2} \right]}$,
where $\mathcal{A}$ is a normalization constant.
The dependence of \eref{def_kwstate} on the measuring apparatus can be then effectively characterized by the coupling-strength parameter $\sigma = \sigma_{A/B}$ and thus \eref{two} reads~\cite{AppC}:
\begin{eqnarray}\label{general}
    \langle y_A(t)y_B(\tau)\rangle_{\sigma_A} = \frac{1}{2} \sum_{i,j} a_i \mathcal{E}_{j,i}\mathcal{B}_{j,i}(\tau)  + c.c,
\end{eqnarray}
where $\mathcal{B}_{j,i}(\tau) = \langle a_{j}|\hat B(\tau)| a_{i}\rangle$ are the matrix elements of the Heisenberg operator $\hat B(\tau) = \hat U_\tau^\dagger \hat B \hat U_\tau$, and we have defined $\mathcal{E}_{j,i} = c_{j}^*\exp{\big[ -{\left( a_i - a_{j} \right)^2}/{4\sigma_A^2} \big]}c_i$.
The expectation value in \eref{general} now bears a subscript $\sigma_A$ that reinforces the idea that this result depends on the measurement scheme. That is, two-time expectation values are generally contextual~\cite{dressel2012contextual}.
Concerning the assumption of the meter wavepacket to be represented by a Gaussian operator, it is shown in Ref.~\cite{AppC} that \eref{general} can be derived for non-specific meter wavefunction shapes.

The result in \eref{general} can be generalized to systems made of $N$ interacting particles. For that, we consider a general (non-separable) state $|\psi (t)\rangle = \sum_{i_1,..,i_N} c_{i_1,..,i_N} | a_{i_1},..,a_{i_N} \rangle$, where $c_{i_1,..,i_N} = \langle a_{i_1},..,a_{i_N} | \psi (t) \rangle$. We define also the many-body intensive operator $\hat{A} = \sum_{\xi=1}^N 
\hat I\otimes \cdots \otimes \hat A_\xi \otimes \cdots \otimes  \hat I/N$, where the index $\xi$ only denotes the degree of freedom that the single-particle operator, $\hat A_\xi$, acts on. Then, the analogous of \eref{general} for a many-body system reads~\cite{AppD}:
\begin{equation}\label{MB} 
 \langle {y_A(t)y_B(\tau)} \rangle_{\sigma_A} = \frac{1}{2N} \! \sum_{\xi=1}^N
 \sum_{\substack{i_1,..,i_N\\j_1,..,j_N}}^\infty \!\!\!\!
 a_{i_\xi} \mathcal{E}_{\substack{i_1,..,i_N\\j_1,..,j_N}}
    \mathcal{B}_{\substack{i_1,..,i_N\\j_1,..,j_N}} \! +  c.c,
\end{equation}
where we have defined the matrix elements:
\begin{subequations}
 \begin{equation}\label{S}
    \mathcal{B}_{\substack{i_1,..,i_N\\j_1,..,j_N}} = \langle a_{j_1},..,a_{j_N}|\hat{B}(\tau) |a_{i_1},..,a_{i_N}\rangle,
 \end{equation}
 \begin{equation}\label{E}
    \mathcal{E}_{\substack{i_1,..,i_N\\j_1,..,j_N}} = c_{j_1,..,j_N}^*  \exp\Bigg[-\frac{\big(\sum_{\nu}^N a_{i_{\nu}}  -  a_{j_{\nu}}\big)^2}{4\sigma_A^2 N^2} \Bigg] c_{i_1,..,i_N}.
 \end{equation}
\end{subequations}
For $N = 1$ \eref{MB} trivially reduces to \eref{general}.
For $N\neq 1$, the backaction of the measurement of A can induce entanglement among particles~\cite{cabrillo1999creation,chou2005measurement}. 

At this point we want to address the question of whether there exists a specific measurement regime where the result in \eref{MB} becomes non-contextual, viz., $\langle y_A y_B \rangle_{\sigma_A} \approx \langle y_A y_B \rangle$. 
For that, we define the effective dimension of the system, $\textrm{d}_{\text{eff}}$, as a measure of the average width of the relevant spectrum of the system with respect to $\hat{A}$, i.e:
   $\textrm{d}_\textrm{eff} := {\sum_{\nu=1}^N \max(\Delta A_\nu)}/{N}$, 
where $\max(\Delta A_\nu)$ is the maximum distance between the occupied upper and lower bounds of the spectrum of $\hat A_\nu$. 
Then, a simple inspection of the matrix elements in \eref{E} shows that for any coupling $\sigma_A$ fulfilling the condition
  $\textrm{d}_\textrm{eff} \ll \sigma_A$,  
one always measures~\cite{AppE}:
\begin{equation}\label{noncontextual} 
 \langle {y_A(t)y_B(\tau)} \rangle = {\langle \psi(t)| \hat{A}(t) \hat{B}(\tau) |\psi(t)\rangle} + c.c.
\end{equation}

The condition $\textrm{d}_\textrm{eff} \ll \sigma_A$ defines what we call the ideally-weak measurement (IWM) regime: the regime where one always measures the same expectation value independently of $\sigma_A$~\cite{misc1}.
This is the case even if the joint probability distribution of measuring $y_A(t)$ and $y_B(\tau)$ depends on $\sigma_A$. 
This result adds to previous findings~\cite{anastopoulos2006classical,di2008weak,dressel2010contextual,dressel2012contextual} by showing that, while quantum backaction is needed for correlation functions to be contextual, viz., $\langle y_A y_B \rangle_\sigma 
 \Rightarrow P(y_{B},y_{A})_{\sigma_A}$,
the contrary is not true for a general class of experiments, viz., $P(y_{B},y_{A})_{\sigma_A}
    \;\;{\nRightarrow} \;\;
    \langle y_A y_B \rangle_{\sigma_A}$.

Yet, the limit $\textrm{d}_\textrm{eff} \ll \sigma_A$ implies a strict cancellation of the sigma-dependence of the joint probability $P(y_A,y_B)_{\sigma_A}$ when integrated over all possible values of $y_A$ for $\sigma_A$s larger than a given threshold $\sigma_{A_o}$~\cite{AppF}, i.e.:
\begin{equation}\label{IWM}
\textrm{IWM:}  \;\;\; \frac{d}{d\sigma_A} \int dy_A  P(y_A,y_B)_{\sigma_{A}} = 0, \; (\forall \sigma_A > \sigma_{A_o}).
\end{equation}
Thus, by assessing the validity of \eref{IWM} for a reasonable number of distinct measurement set-ups (with different system-meter coupling-strenghts), an experimentalist can assert whether or not he/she is working in the IWM and hence whether the measurements conducted in the laboratory are generalized von Neumann measurements of the type described here. 

Making sure that one is operating in the IWM regime, however, does not guarantee that the measurement of A is non-invasive.
This is a crucial point that can be appreciated by rewriting the final state of the system in \eref{def_kwstate} using a first order Taylor expansion of $ \Omega_{y_A-a_i}$ and $\Omega_{y_B-b_j}$ around $y_A$ and $y_B$ in the limit of $\sigma_{A/B}\to\infty$~\cite{AppG}:
\begin{eqnarray}
|\psi_{A,B}\rangle = \Omega_{y_B}\Omega_{y_A}  \Big(\mathbb{I} \;  \text{-} \; \frac{y_A}{\sigma_A^2}\hat{B}  \Big)  \Big(|\psi(\tau)\rangle \; \text{-} \;   \frac{y_B}{\sigma_B^2} |\widetilde{\psi}(\tau)\rangle \Big),
\label{perturbation}
\end{eqnarray}
where we have defined $|\psi(\tau)\rangle = \hat U_\tau |\psi(t)\rangle$ and $|\widetilde{\psi}(\tau)\rangle = \hat U_\tau \hat{A} |\psi(t)\rangle$.
Expression \eqref{perturbation} tells us that the state of the system right after two ideally-weak measurements can be written as a superposition of two states, and that only the first one, i.e., $\Omega_{y_B}\Omega_{y_A}  \left(\mathbb{I} \; \text{-} \; \frac{y_A}{\sigma_A^2}\hat{B}  \right) |\psi(\tau)\rangle$, contains information about the system having evolved freely from $t$ to $\tau$. 
Generally, the second term in \eref{perturbation} is not proportional to $\hat B\hat U_\tau |\psi(t)\rangle$ and hence it represents the non-negligible backaction of the first measurement on the subsequent evolution of the system. 

Only when the state of the system $|\psi(t)\rangle$ can be approximated by an eigenstate of the operator $\hat{A}$, i.e.: $\hat{A} |\psi(t)\rangle \approx \langle \hat{A} \rangle |\psi(t)\rangle$,
then the backaction of the first measurement is avoided, and hence property A is said to be macrorealistic.
In short, \eref{noncontextual} reduces to
$\langle y_A(t)y_B(\tau)\rangle = \langle y_A(t) \rangle  \langle y_B(\tau) \rangle$,
which is the definition of macrorealism for a pure state, i.e., $P(y_{B},y_{A})_{\sigma_A} = P(y_A)_{\sigma_A}P(y_B)_{\sigma_A}$.

Let us recapitulate. While an IWM of a non-macrorealistic property does induce a backaction on the system (see \eref{perturbation}), the resulting effects at the level of the reduced probabilities $\int dy_A  P(y_A,y_B)_{\sigma_{A}}$ in \eref{IWM} are independent of the properties of the measuring apparatus.
This is a very interesting result, valid also for general mixed states, that can be exploited to verify that a given experimental set-up can be effectively represented by a generalized von Neumann measurement model.
This is precisely the type of ``good'' measuring apparatus that, as shown in the above paragraph, happen to be non-invasive for macrorealistic properties. Therefore, as it will be shown in the following, the use of the IWM conditions in combination with a given test of macrorealism can be used to close the clumsiness loophole.

For general (mixed) states, macrorealism can be defined as~\cite{kofler2013condition}:
\begin{equation}\label{MR_mixed}
    \textrm{MR:}\;\;\;P(y_A,y_B) = \sum_\lambda \rho(\lambda) P_\lambda(y_A)P_\lambda(y_B),
\end{equation}
where $\lambda$ specifies all properties of the system. 
Due to the mixedness of the initial state, the violation of macrorealism can be hidden in the statistics of the experiment. A test of macrorealism can then be based on the statistical version of the ``non-invasive measurability'' condition, also referred to as ``no-signaling in time'' (NSIT)~\cite{kofler2013condition}:
\begin{equation}\label{NSIT}
   \textrm{NSIT:}\;\;\; P(y_B) = \int dy_A P(y_A,y_B).
\end{equation}
This condition, originally proposed as an alternative characterization of macrorealism, differs from the Leggett-Garg inequalities~\cite{kofler2013condition,clemente2015necessary}.
Hoerver, while MR $\Rightarrow$ NSIT, the violation of NSIT can only indicate either (i) that the system is non-macrorealistic or (ii) that the system is macrorealistic but subjected to a measurement technique that happens to disturb the system~\cite{misc5}. To discard (ii) above, we propose the following:
\begin{enumerate}[(S1)]
    \item Make sure that the measurement of A at time $t$ is carried out in the IWM regime by testing \eref{IWM} for a reasonable number of measurement set-ups $\{\sigma_A\}$.
    
    \item Equate the resulting reduced probability distributions as in \eref{NSIT}. Property A is macrorealistic if NSIT is fulfilled and non-macrorealistic otherwise.
\end{enumerate}
Note that NSIT $\Rightarrow$ IWM, and therefore under the fulfillment of \eref{NSIT} the condition \eref{IWM} is trivially fulfilled. Whenever NSIT is violated, however, being under the IWM regime will be the only warranty that the experimental set-up represents a ``good'' measuring apparatus (i.e. non-invasive for macrorealistic properties).

Assessing the IWM condition in \eref{IWM} requires to design a number of different measurement set-ups. The larger the number of measurement set-ups that are compared one to each other, the more trustworthy the test of macrorealism will be. Put differently, the probability that \eref{IWM} is fulfilled simultaneously by a number of classically invasive measurement apparatuses (different from the generalized von Neumann measurements described here) decreases with the number of experimental set-ups itself. Escaping this test would simply be too conspiratorial a loophole to take seriously.

Let us mention that the protocol described by (S1) and (S2) only assesses macrorealism at time $t$ and with respect to an intensive property A. 
In a test of genuine macrorealism the validity of \eref{NSIT} should be proven for any observable at any time. 
This is obviously a prohibitive experimental task, and hence it is common to associate macrorealism only to a given observable of interest~\cite{palacios2010,goggin2011violation,knee2012violation,athalye2011investigation}. 
Anyhow, genuine macrorealism is not expected in general, not at least for operators representing extensive properties such as, e.g., the angular momentum or magnetization. 
Yet, examples of macrorealism for general intensive properties of the type considered here, far from being atypical, can be common for large systems made of weakly-interacting particles.
Consider, e.g., a system defined by separable wavefunctions $|\psi (t)\rangle=|\psi_1 (t)\rangle \otimes ... \otimes |\psi_N (t)\rangle$ where $|\psi_i(t)\rangle$ are all identical single-particle states.
To determine whether the state $|\psi(t) \rangle$ is an eigenstate of an intensive property $\hat{A}$, i.e., $\hat{A} |\psi(t) \rangle \approx \langle \hat{A} \rangle |\psi(t) \rangle$ with $\langle \hat{A} \rangle = \sum_{\xi}^N \langle \psi_\xi(t) | \hat A_\xi | \psi_\xi(t) \rangle/N$, we check the soundness of the identity $\langle  \hat{A}^2 \text{-} \langle \hat{A}\rangle^2 \rangle=0$.
By writing $\hat{A}^2 = N^{\text{-}2} \sum_{\xi}^N \left(  \hat A_\xi \hat A_\xi+  \sum_{\nu \neq \xi}^N \hat A_\xi \hat A_\nu \right)$, it is easy to realize that the expectation value $\langle \psi(t) |\hat{A}^2 |\psi(t)\rangle$ reads:
$\langle \hat{A}^2(t) \rangle = N^{\text{-}2} \sum_{\xi=1}^N \Big[ \langle \hat A_\xi^2(t) \rangle 
 + \sum_{\nu\neq \xi}^N \langle \hat A_\xi(t) \rangle \langle \hat A_\nu(t) \rangle \Big]$.
Therefore, in the limit $N\to\infty$ we get $\langle \hat{A}^2 \rangle = \langle \hat{A} \rangle^2$, so we conclude that $\hat{A} |\psi(t)\rangle = \langle \hat{A} \rangle |\psi(t)\rangle$. That is, even if individually $|\psi_\xi(t)\rangle$ are not eigenstates of $\hat A_\xi$,
in the limit $N\to\infty$ one could arguably speak of  macrorealism of any intensive property A~\cite{misc2}. This is in contrast with the quantumness of the system itself, which, being preserved, would prevent us to talk about realism at the microscopic level~\cite{emary2013}.

\begin{figure}
  \hspace{0.6cm}\includegraphics[width=0.85\columnwidth]{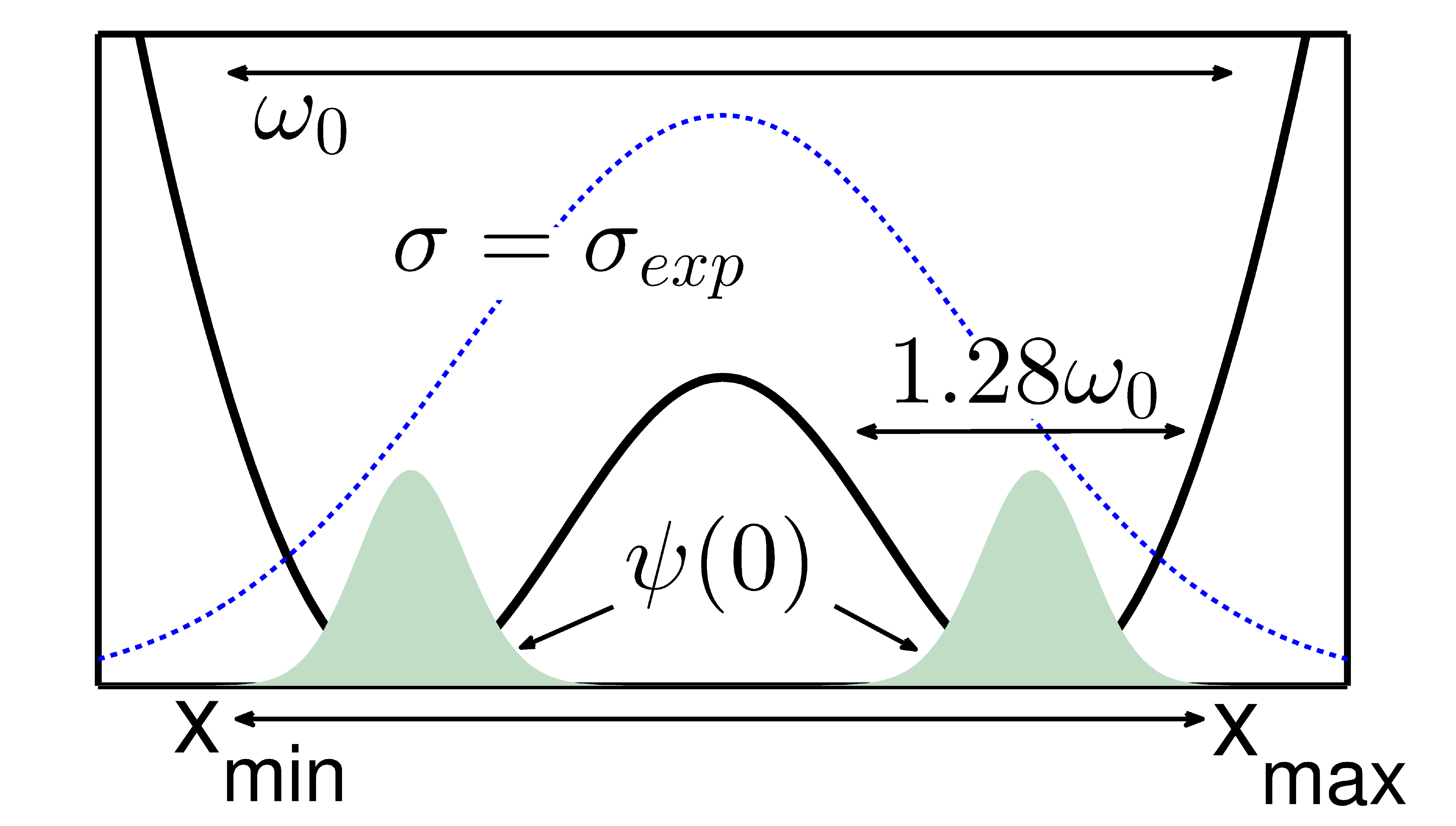}
  \includegraphics[width=\columnwidth]{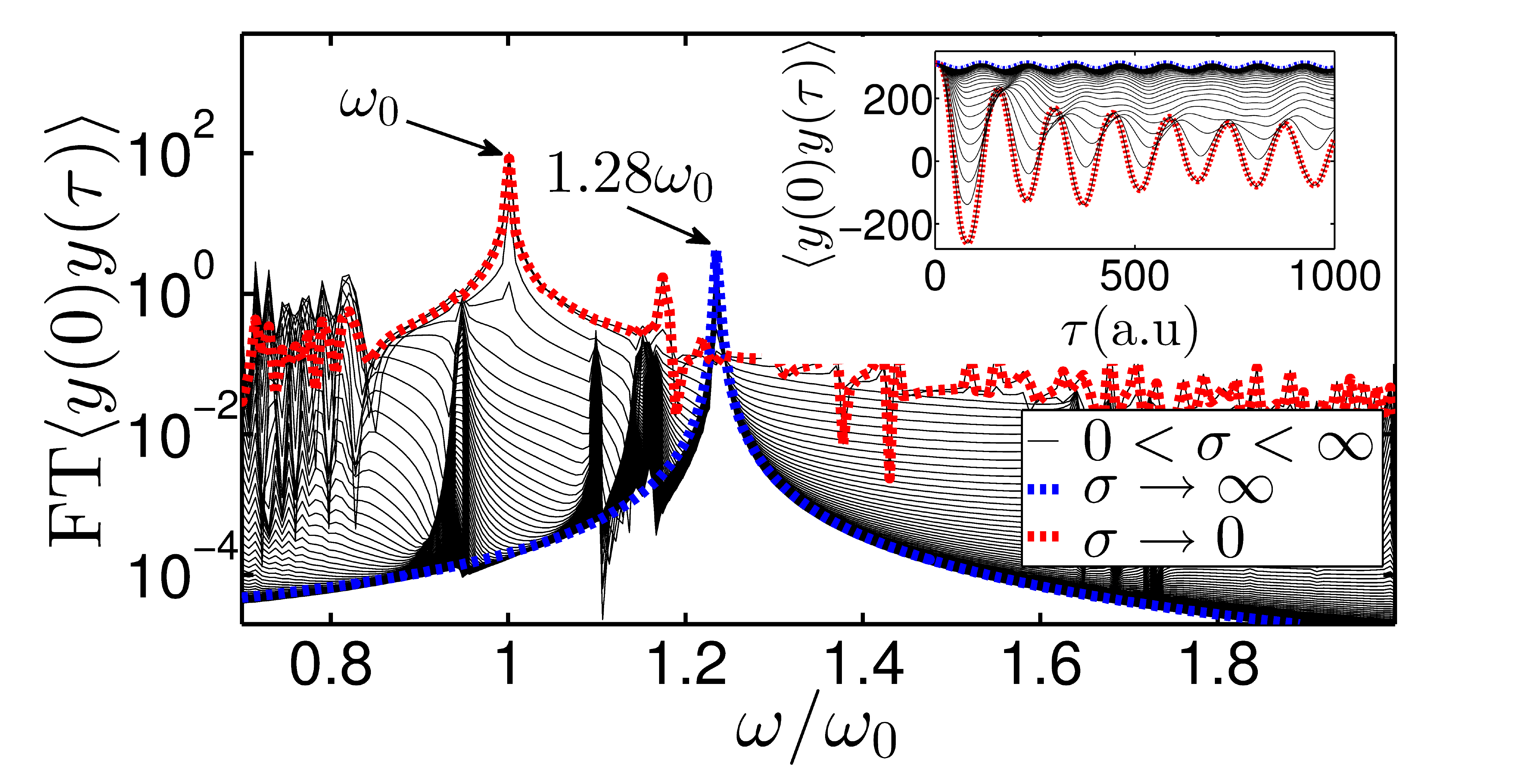}
  \caption{Top panel: schematic picture of the double-well oscillator. The potential energy curve is plot in solid black line.
  The initial state of the system (area in green) is taken to be the ground state of the system. Two main frequencies are involved in the dynamics of the system, viz., $\omega_0$ and $1.28\omega_0$, related respectively with the inter-well and intra-well dynamics.
  The relevant upper and lower bounds of the spectrum of $\hat X$ are denoted by $\textrm{x}_\textrm{max}$ and $\textrm{x}_\textrm{min}$ respectively, and the exponential function defined in \eref{E} is depicted for a particular value of $\sigma_X$ in dashed blue line.
  Bottom panel: semi-log plot of the Fourier transform of the autocorrelation function in \eref{general} as a function of $\sigma_X$ and $\tau$ (solid black lines). The limits of $\sigma_X\to 0$ and $\sigma_X\to\infty$ are shown respectively in dashed red and blue lines. In the inset: the same results but for the autocorrelation function.}
  \label{plot1}
\end{figure}

To illustrate the proposed test of macrorealism, we consider a simple numerical experiment. We will evaluate the autocorrelation function of the center-of-mass position operator, $\hat{{X}} = \sum_{\xi}^N \hat X_{\xi}/N$, for a number $N$ of uncoupled one-dimensional double-well oscillator (see the top panel of \fref{plot1}). Hereafter we use atomic units, $\hbar = m = 1$, and define the single-particle oscillator's Hamiltonian as $\hat H = {\hat P^2}/{2} + {\omega_0^2\hat X^2}/{2} + \cosh^{-2}{\alpha\hat X}$, where $\hat P$ is the momentum operator, and the natural frequency of the underlying harmonic oscillator is $\omega_0 = 4.3\cdot 10^{-3}$a.u. The characteristic width of the barrier between the two wells is set to $\alpha=5\cdot 10^{-2}a.u$. We choose $t=0$ such that the only relevant time in the discussion is $\tau$.
We consider that the oscillators are all initially prepared in the ground state. 
Then, by taking the non-interacting limit of \eref{MB}, we find (for arbitrary initial conditions see~\cite{AppH}):
\begin{eqnarray}\label{collective}
    \langle {y_Ay_B} \rangle_\sigma = \frac{1}{2N} \sum_{i,j}^\infty  \mathcal{E}_{j,i} \mathcal{B}_{j,i} 
     \big(  a_{i} \!+\! (N \text{-} 1) \langle  \hat A(t) \rangle   \big) + c.c,
\end{eqnarray}
which in the limit of $N\to \infty$ reduces to $\langle{y_A(t)}\rangle \langle y_B(\tau) \rangle$.

The dynamics of a single oscillator for different values of $\sigma_X$ is shown in \fref{plot1}.
For a projective measurement, i.e., $\sigma_X\to 0$, the dynamics presents a central resonance peak at $\omega_0$ (in dashed red line). This is due to the strong perturbation induced by the projective measurement at $t=0$, which yields a subsequent dynamics characterized by a large amplitude (over-the-barrier) oscillation. Contrarily, in the limit $\sigma_X\to \infty$ the measurement produces only a small perturbation to the initial state and yields an ensuing dynamics confined in the wells with a characteristic frequency $\omega = 1.28\omega_0$ (in dashed blue line). In between these two regimes, an infinite number of dynamics can be inferred depending on the system-meter coupling strength (in black solid lines).

\begin{figure}
  \includegraphics[width=\columnwidth]{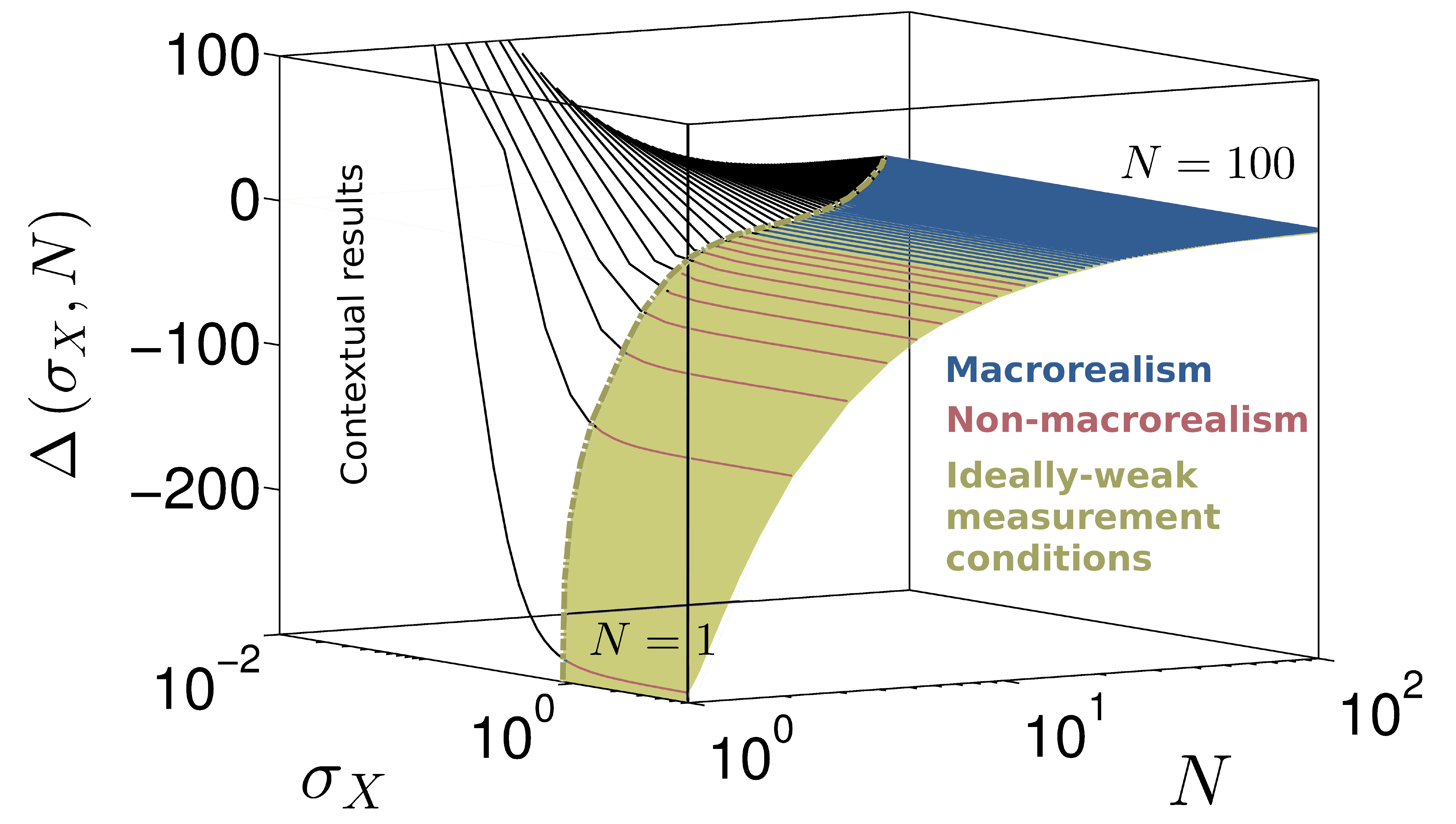}
  \caption{$\Delta (\sigma_X,N)$ as a function of $\sigma_X$ and the number $N$ of oscillators for $\tau=33.3\pi$. Non-contextual results are differentiated from contextual results by an underlying yellow surface. Non-macrorealism and macrorealism results are shown in red and blue respectively.}
  \label{plot2}
\end{figure}

To conclude whether the position of a single oscillator is macrorealistic, we first need to ensure that the measurement of $X$ at time $t=0$ is carried out in the IWM regime (i.e., S1), and then compare the expectation values $\langle y(0)y(\tau) \rangle$ and $\langle y(0) \rangle \langle y(\tau) \rangle$ (i.e., S2). Note that since our example only considers pure states, the condition in \eref{MR_mixed} can be replaced by the simpler one $P(y_{B},y_{A}) = P(y_A)P(y_B)$.
We address (S1) and (S2) in a compact way using the quantity
\begin{equation}
    \Delta (\sigma_X,N) = \frac{d\langle y_X(0)y_X(\tau) \rangle}{d\sigma_X}d\sigma_X - \Delta_{QC}
\end{equation}
where $\Delta_{QC} = \langle y_X(0)y_X(\tau) \rangle - \langle y_X(0) \rangle \langle y_X(\tau) \rangle$.
Whenever $\Delta (\sigma_X,N)$ becomes constant, \eref{IWM} is fulfilled, and whether the center-of-mass position is macrorealistic or not can be checked by simply assessing $\Delta (\sigma_X,N)$ in the asymptotic region. That is, X is macrorealistic if $\Delta (\sigma_X,N)$ vanishes in the asymptotic region and non-macrorealistic otherwise. 
In Fig. \ref{plot2} we plot the quantity $\Delta (\sigma_X,N)$ as a function of $\sigma_X$ and the number $N$ of oscillators. A single oscillator is non-macrorealistic as $\Delta (\sigma_X,1)$ asymptotically converges to a non-zero value.
For a large enough number of oscillators, however, the dynamics of $\hat{X}$ becomes independent of $\sigma_X$ which is a clear signature of macrorealism as defined in \eref{MR_mixed}. 
In general, the $N$ oscillators become entangled right after the first measurement process and this allows a smooth transition (exponential decay with $N$) between the non-macroreaslitic and macrorealistic results. 

\textit{Conclusion.---} 
Quantum dynamics is ambiguous unless it goes along with a proper discussion of the system-meter interaction. This applies also to the great majority of tests of macrorealism, where contextuality appears in the form of a clumsiness loophole. 

In this Letter we have proven a sufficient condition for the non-contextuality of reduced one-time probability densities for a family of, classically non-perturbative, generalized von Neumann measurements. 
This condition, named IWM regime, can be assessed according to \eref{IWM}, which in turn requires to design a number of different experimental set-ups.
For a large enough sample of set-ups, probably implemented at different laboratories, falsifying \eref{IWM} would require a loophole too conspiratorial to be taken seriously.

Based on this result we have proposed a test of macrorealism that consists on witnessing the so-called no-signaling in time condition, \eref{NSIT}, under the fulfillment of the IWM regime, \eref{IWM}. The resulting protocol allows for tests in situations (e.g., unbounded and non-dichotomic properties) where Leggett-Garg inequalities and ideal negative measurement cannot be used at all.


\section{Acknowledgements}
D.P. and X.O acknowledge funding from Fondo Europeo de Desarrollo Regional (FEDER), the Ministerio de Ciencia e Innovaci\'{o}n through the Spanish Project TEC2015-67462-C2-1-R, the Generalitat de Catalunya (2014 SGR-384), the European Union's Horizon 2020 research and innovation program under grant agreement No Graphene  Core2 785219 and under the Marie Sklodowska-Curie grant agreement No 765426 (TeraApps).
G.A. acknowledges financial support from the European  Union’s  Horizon 2020 research and innovation programme under the Marie Sklodowska-Curie grant agreement No 752822, the Spanish Ministerio de Economía y Competitividad (CTQ2016-76423-P), and the Generalitat de Catalunya (2017 SGR 348).

 \bibliographystyle{apsrev4-1}
 \bibliography{bibliography}

\begin{thebibliography}{53}
\bibitem [{mis({\natexlab{a}})}]{misc3}%
  \BibitemOpen
  \href@noop {} {} ({\natexlab{a}}),\ \bibinfo {note} {sums should be replaced
  by integrals when the observable $A$ has a continuum spectrum. Note also that
  this scheme defines a positive operator value
  measure~\cite{kofman2012nonperturbative,tamir2013introduction}.}\BibitemShut
  {Stop}%
\bibitem [{mis({\natexlab{b}})}]{misc1}%
  \BibitemOpen
  \href@noop {} {} ({\natexlab{b}}),\ \bibinfo {note} {let us also remark that
  in order to extend the above result to multiple-time (more than two)
  measurements, i.e., $\langle {y_Ay_B...y_Z} \rangle$, the condition
  $\textrm{d}_\textrm{eff} \ll \sigma_A$ must be fulfilled by all but the last
  subsequent measurements, $\sigma_A$,$\sigma_B$,...,$\sigma_Y$.}\BibitemShut
  {Stop}%
\bibitem [{mis({\natexlab{c}})}]{misc5}%
  \BibitemOpen
  \href@noop {} {} ({\natexlab{c}}),\ \bibinfo {note} {as we have discussed in
  relation with \eref{perturbation}, performing an ideally-weak measurement
  does not guarantee that quantum backaction is evaded. Therefore, IWM
  $\Rightarrow$ NSIT $\Rightarrow$ MR is true only if all pure states
  conforming a given mixed state are eigenstates of property A.}\BibitemShut
  {Stop}%
\end{thebibliography}

 \end{document}